# Deep Charge: A Deep Learning Model of Electron Density from One-Shot Density Functional Theory Calculation

Taoyuze Lv,[1] Zhicheng Zhong,[2,*] Yuhang Liang,[3] Feng Li,[1] Jun Huang,[3] Rongkun Zheng[1,†]

[1] School of Physics, The University of Sydney, NSW 2006, Australia

[2] Ningbo Institute of Materials Technology and Engineering, Chinese Academy of Science, Ningbo, Zhejiang, 315201, China

[3] School of Chemical and Biomolecular Engineering, The University of Sydney, NSW 2006, Australia

* zhong@nimte.ac.cn

† rongkun.zheng@sydney.edu.au

## Abstract

Electron charge density is a fundamental physical quantity, determining various properties of matter. In this study, we have proposed a deep-learning model for accurate charge density prediction. Our model naturally preserves physical symmetries and can be effectively trained from one-shot density functional theory calculation toward high accuracy. It captures detailed atomic environment information, ensuring accurate predictions of charge density across bulk, surface, molecules, and amorphous structures. This implementation exhibits excellent scalability and provides efficient analyses of material properties in large-scale condensed matter systems.

## Introduction

In the vast domain of materials science and condensed matter physics, electron charge density is a fundamental property that describes the spatial distribution of electrons in a material. It provides invaluable insights into overall material behaviors and is an essential quantity in various quantum mechanical frameworks, especially in density functional theory (DFT) [1], where the ground-state properties of a many-body system can be described as functionals of the electron charge density. Commonly, DFT solves the Kohn-Sham (KS) equations[2] via the self-consistent loop[3] to adopt the ground state charge density $\varrho(r, \{R_i\}_{i=1}^n)$, where charge density $\varrho$ is a function of the space position $r$ and the positions of all $n$ atoms in the system $\{R_i\}_{i=1}^n$. A significant challenge of this approach is the computational complexity of $O(N^3)$ with respect to the number of electrons. The harshness of density functional optimization impedes the application of DFT in mesoscopic systems with thousands or even millions of atoms. This limitation has spurred the search for alternative approaches that can efficiently represent charge density without compromising accuracy.

The electron density generally obtains "nearsightedness" and screening effect in many-atom systems[4,5]. These effects make it possible to introduce a radius $r_c$ around a space point, beyond which any perturbation on the atomic environment only provides negligible impacts on the electron density at that point. This is the locality property of electron density, which does not apply to the KS wavefunctions. In that case, the original time-consuming KS self-consistent calculation can be turned into mapping local atomic environment information to the electron density. Such a mapping can naturally be obtained via techniques such as machine learning. A well-trained machine learning model can be used for predicting charge density in different atomic systems, and even be scalable for determining large-scale material systems. Currently, several machine-learning models have been proposed to predict electron charge density as alternatives to first-principles calculations, as listed in **TABLE 1**.

One commonly used strategy is to express the charge density as the summation of a set of basis functions such as spherical harmonics, with the coefficients fitted via machine learning[6-12]. The rotational invariance of charge density can be automatically protected with a reasonable basis set. These methods are efficient but mainly used in charge-localized systems, and the accuracy is highly dependent on the quality and complexity of basis functions and usually requires DFT calculations of plenty of different atomic structures to train. On the other hand, the grid-based methods can be adopted to obtain more flexibility[13-18]. These methods focus on mapping every unique local environment around real space points to non-degenerate values in the latent space, which thus contains enough information for charge density representation. The mapping should generally be invariant under three physical symmetry operations: translation, rotation, and permutation, as will be discussed in detail below. Achieving these without losing structural detail information is a nontrivial task. Moreover, grid point methods usually require large numbers of data points from various structures to obtain satisfactory accuracy.

Herein, to overcome the above challenges for the grid-based method, we introduce a deep-learning charge model (Deep Charge) to represent electron charge density. We have adopted the embedding framework from Deep Potential (DP)[19] to extract the local atomic environment, which naturally preserves the system's symmetry. This implementation requires very few structures for training. Remarkably, a one-shot DFT calculation is sufficient for simple systems like crystalline silicon and aluminium to reach a mean absolute error (MAE) below $7\times10^{-4}$ e/Å$^3$. The model provides a scalable representation of charge density in bulk, surface, and complex systems such as alloy and amorphous, involving semiconducting and metallic systems. Our Deep Charge model can thus efficiently assist in the prediction and analysis of charge density in large-scale condensed matter systems with DFT accuracy.

**TABLE 1.** The published works on machine-learning charge density. The top 7 methods belong to basis function methods, while the bottom 6 are grid-point methods.

| Method | Systems | Accuracy | Data size | Ref. |
|---|---|---|---|---|
| SA-GPR | Small molecules | MAE: 1.0% ~ 1.2% | ~1000 structures | [6] |
| SA-GPR | Organic dimers | rMAE[b]: 0.3%~1.8% | ~2300 structures | [7] |
| SALTED | Al, Si, I$_h$-Ice | %RMSE[c]: ~1.5% | 100 structures | [8] |
| EGNN[a] | DNA | rMAE ~1% | 922 structures | [9] |
| SchNOrb | Small molecules | \ | 25000 structures | [10] |
| EGNN | Water clusters | rMAE: ~1.4% | 6000 structures | [11] |
| ML-HK map | Small molecules | \ | 100~350 structures | [12] |
| Gaussians | PE and Al thin films | RMSE[d]: ~5×10$^{-4}$ | > 10$^7$ points | [13] |
| CGCNN | Polymer, zeolite | RMSE: 0.04~0.18 | >3×10$^5$ points | [14] |
| Deep Density | Molecules, Al | rRMSE[e]: 0.5%~2.2% | 10$^5$~10$^6$ points | [15] |
| Gaussians | Organics | RMSE: 0.03~1.52 | ~1.5×10$^9$ points | [16] |
| EGNN | Molecules, cathode | rMAE: 0.06%~0.27% | 10$^3$~10$^5$ structures | [17] |
| JLCDMs | Benzene, Al, MoS$_2$ | RMSE: 0.006~0.008 | ~1.6×10$^6$ points | [18] |
| **This work** | Si, Al, Al-Mg, water | MAE: 0.05%~0.4% | >10$^5$ points | |

[a] Equivariant graph neural network.

[b] Relative mean absolute error. $\sum_i^N |\rho_i - \hat{\rho}_i|/\sum_i^N |\rho_i|$, in which $\rho_i$ is the real charge density of ith grid point, $\hat{\rho}_i$ is the represented value of that point.
[c] Percentage root means square error, defined in the original article.
[d] Root mean square error. $\sqrt{\sum_i^N (\rho_i - \hat{\rho}_i)^2 / N}$, in which N is the total number of grid points.
[e] Relative root means square error. $\sqrt{\sum_i^N (\rho_i - \hat{\rho}_i)^2} / \sqrt{\sum_i^N \rho_i^2}$.

# Deep Charge model

The continuous charge density is discretized into a space grid in grid point methods. Suppose there are *n* atoms in the system, the charge density value at position **r** can be written as a function of the coordinates of atoms $\{R_i\}_{i=1}^n$,

$$\rho = \varrho(r, \{R_i\}_{i=1}^n) \tag{1}$$

where all the information of the system is required to determine the density at a single point. Instead, we can use only the coordinates of neighboring atoms $\{R_i | i \in n_{r_c}(r)\}$, where $n_{r_c}(r)$ denotes the set of atoms within the cutoff radius $r_c$ surrounding **r** with a size of $n_c$.

$$\varrho(r, \{R_i\}_{i=1}^n) \cong \varrho(r, \{R_i | i \in n_{r_c}(r)\}). \tag{2}$$

Ideally, this charge density representation should be designed carefully to fulfill the invariance under three symmetry operations:

1. Any translation $\hat{\mathcal{T}}_d$ of the whole system $\hat{\mathcal{T}}_d \varrho(r, \{R_i\}_{i=1}^n) = \varrho(r + d, \{R_i + d\}_{i=1}^n)$ should result in the same value as $\varrho(r, \{R_i\}_{i=1}^n)$;
2. Any operation $\hat{O} \in O(3)$ centered at **r** should obey $\hat{O}\varrho(r, \{R_i\}_{i=1}^n) = \varrho(r, \{R_i\}_{i=1}^n)$;
3. Any permutation $\hat{\pi}$ on labels of the same atomic species fulfills $\varrho\left(r, \{R_{\hat{\pi}(i)}\}_{i=1}^n\right) = \varrho(r, \{R_i\}_{i=1}^n)$.

To preserve the translational invariance of the density, the relative coordinates $r_i = r - R_i = (x_i, y_i, z_i)$ are adopted:

$$\varrho(r, \{R_i | i \in n_{r_c}(r)\}) = \varrho(\{r_i | i \in n_{r_c}(r)\}). \tag{3}$$

The density can be represented via neural networks as

$$\varrho(\{r_i | i \in n_{r_c}(r)\}) = \mathcal{F}(\mathcal{D}(\{r_i | i \in n_{r_c}(r)\}; \theta_d); \theta_f), \tag{4}$$

where $\mathcal{D}$ represents the descriptor that extracts all the local environment information, and $\mathcal{F}$ represents the fitting net for charge density. The trainable neural network parameters for them are $\theta_d$ and $\theta_f$, respectively. In the implementation of Deep Charge, the descriptor is constructed to conserve the rotation and permutation symmetry. Herein, we use the two-body embedding DeepPot-SE descriptor with radial and angular information[20]:

$$\mathcal{D} = \frac{1}{n_c^2} \mathcal{G}^T \mathcal{R} \mathcal{R}^T \mathcal{G} \tag{5}$$

in which $\mathcal{R} \in \mathbb{R}^{n_c \times \{1,4\}}$ is the coordinate matrix with the form of

$$\mathcal{R}_i = s(r_i)\left(1, \frac{x_i}{r_i}, \frac{y_i}{r_i}, \frac{z_i}{r_i}\right) \tag{6}$$

in which $r_i = \|\mathbf{r}_i\|$ is the norm of relative coordinates. $s(r)$ is the switching function defined with a fifth-order polynomial:

$$s(r) = \begin{cases} \dfrac{1}{r}, & r < r_s \\ \dfrac{[u^3(-6u^2 + 15u - 10) + 1]}{r}, & r_s \leq r < r_c \\ 0, & r \geq r_c \end{cases}, \quad u = \frac{r - r_s}{r_c - r_s}, \tag{7}$$

where $r_s$ is the radius where smoothing starts. The construction of $\mathcal{R}\mathcal{R}^T$ is naturally invariant under O(3) symmetry operation. The embedding matrix $\mathcal{G}$ is for obtaining the permutation invariance according to ref.[21]. The form of $\mathcal{G}$ is trained with neural networks $\mathcal{N}$:

$$\mathcal{G}_i = \mathcal{N}(s(r_i)). \tag{8}$$

The fitting net $\mathcal{F}$ obtains a simple fully connected feed-forward network structure. The detailed charge density prediction workflow is shown in **FIG. 1**. The continuous space is discretized into grids. The model will take the atomic configuration and grid points as input and find the local environment for each grid point in parallel, which will be used in the embedding networks to construct the descriptor and then input to the fitting networks. The final charge density can thus be collected from the grid points.

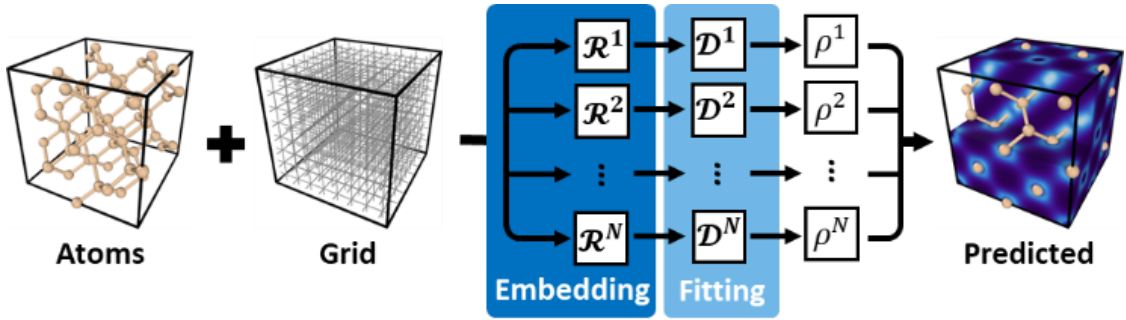

**FIG. 1**. Workflow of Deep Charge prediction. Suppose there are $N$ grid points in the system, each grid point goes through one network separately.

## Results and discussion

### Performance in bulk systems

In this section, we examined the convergence relation between the number of data points for training and the accuracy in bulks. We chose crystalline silicon with 64 atoms for this study. Silicon is an extensively studied semiconductor with localized valence electrons at bond positions. The training sets were based on the charge density obtained from only one structure, while the testing was performed on all the grid points from an unseen structure. The results are shown in **FIG. 2**. In addition, we compared the convergence with models trained on 5 different structures, $10^5$ data points for each structure selected, displayed with blue dashed lines in **FIG. 2(a)**. The model can reach an MAE of $6.7 \times 10^{-4}$ e/Å$^3$ trained with $10^5$ data points. It is noticeable that the one-structure models exhibit similar accuracy to the five-structure models, indicating that a little charge density information of one single structure of structure is enough to learn the patterns in crystals. The parity plot of **FIG. 2(b)** has a coefficient of determination ($R^2$) of 0.999957, also indicating a nearly linear relationship. Meanwhile, as shown in **FIGs. 2(c)-(f)**, the absolute charge density errors are larger around the center of atoms, where the fluctuation and charge density are also higher. The absolute error distribution is also shown in **FIG. S1(a)** in Supplementary Materials (SM)[22].

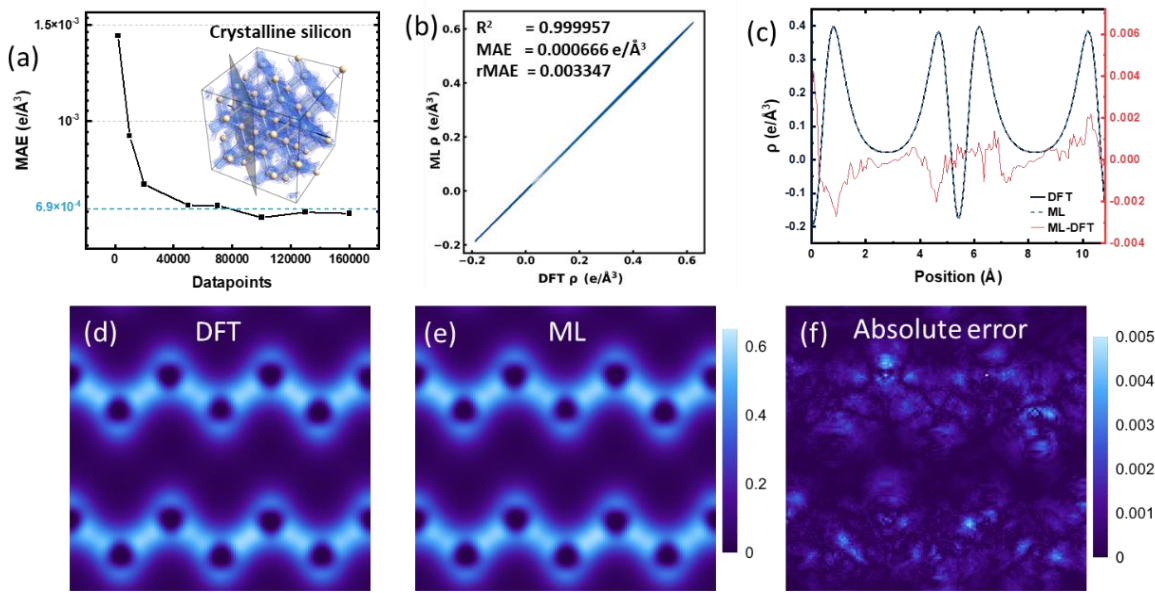

**FIG. 2.** The performance of the Deep Charge model on bulk silicon. (a) The convergence relation between the number of data points used for training and the MAE values for the testing set. (b) The scatter plot of predicted and test densities from a whole testing structure. Notice that the negative values of charge density are induced by the on-site correction terms in projector augmented wave (PAW) pseudopotentials. (c) The charge density along the line indicated in the inset of (a). (d) and (e) is the charge density computed from DFT and machine learning, respectively. The slice is denoted in the inset of (a). (f) The absolute error. Color bars are in the unit of $e/Å^3$.

## Performance in surface systems

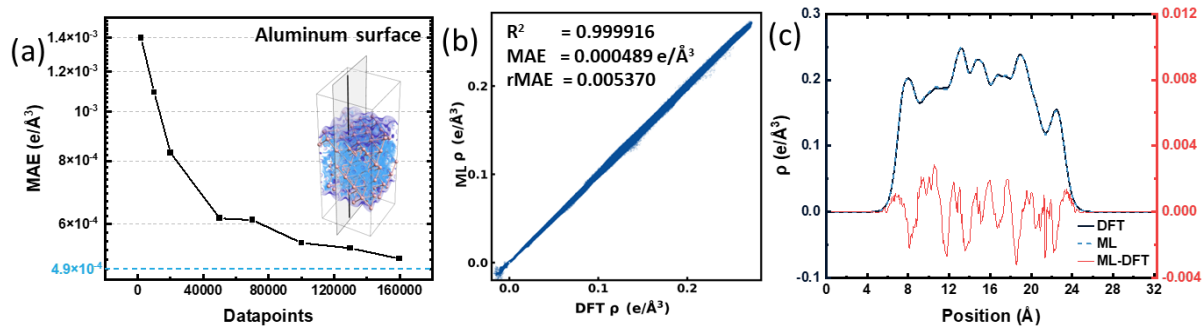

**FIG. 3**. (a) The convergence relation of machine learning model in aluminium surface. (b) the scatter plot of predicted and test densities. (c) The charge density along the line indicated in the inset of (a).

To test the capability of the model under different atomic environments, we further evaluated the charge density representation in surface structures. The aluminium surface with 144 atoms was chosen, because it is a benchmark material for testing charge density representation with highly delocalized valence electrons[8,13,15,18]. As shown in **FIG. 3(a) and (b)**, the accuracy of our model approached below $6\times10^{-4}$ e/Å$^3$ after training with $10^5$ data points, which is quite similar to the five-structure model. We have extracted the charge density along the line across the surface and bond positions, as shown in **FIG. 3(c)**. The charge density near the surface is lower than that in the bulk. This is owing to the surface relaxation, the atoms near the surface are loosely bonded, reducing the charge density between atoms. Moreover, the charge density inside the film also varies considerably with thermal fluctuations than that in silicon, which is also clearly shown in **FIG. 4**. The results from the Deep Charge model greatly overlap that from the DFT calculation, meaning that the embedding approach can accurately capture the subtle changes in the environment, with the error fluctuating under 0.004 e/Å$^3$ along the line.

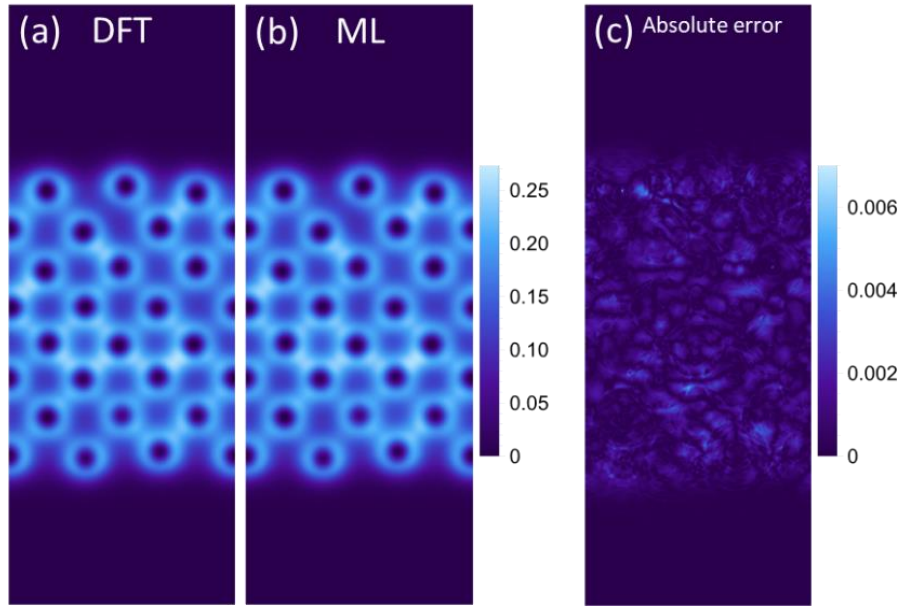

**FIG. 4**. (a) and (b) are the computed charge density from DFT and machine learning, respectively. The slice is denoted in **FIG. 3(a)**. (c) the absolute error.

**Performance and scalability in complex structures**
In this section, we examined the scalability of the machine learning model in complex structures. Scalability is one of the most valuable aims of machine learning, since complex structures normally require a large system size in practical applications, while the ordinary DFT calculations for large supercells are extremely challenging. Therefore, we have studied the model in alloy and amorphous materials. Specifically, aluminium-magnesium alloy (108 atoms), amorphous silicon (64 atoms), and water (48 atoms) were tested.

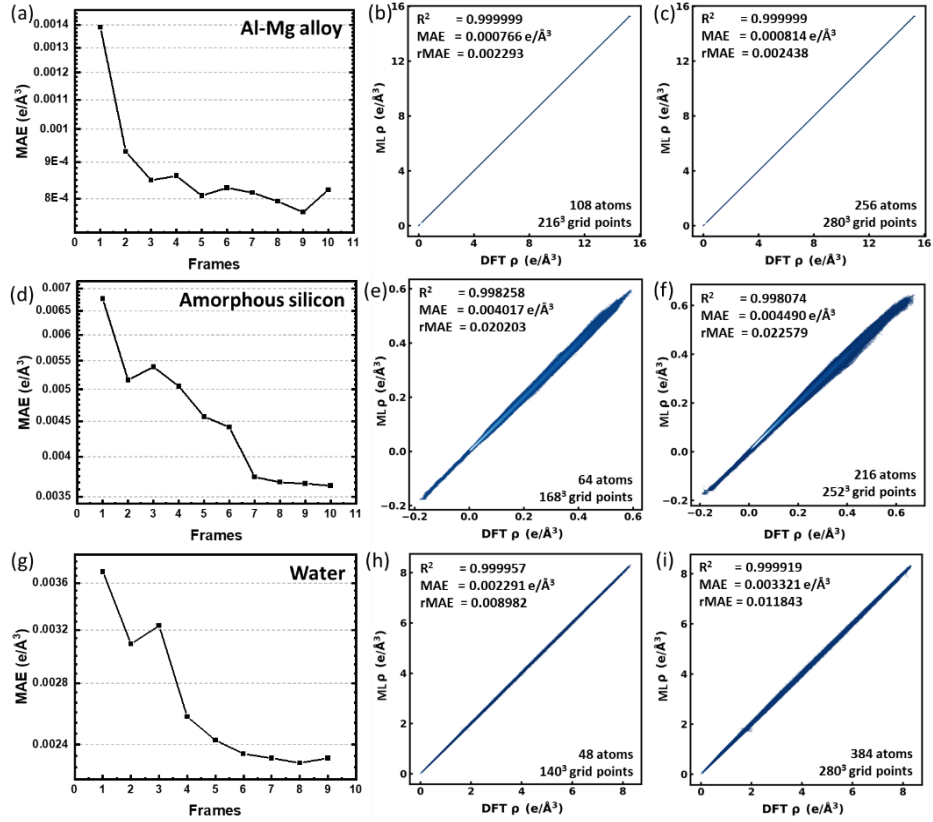

**FIG. 5**. (a), (d) and (g) are the convergence relation between the number of structures used for training and the MAE values for the testing set of aluminium-magnesium alloy, amorphous silicon, and water, respectively. (b), (e) and (h) are the corresponding scatter plots for charge density calculated by DFT and Deep Charge for a whole grid of data from a testing structure. (c), (f) and (i) are the corresponding parity plots in larger supercells.

First, a face-centered cubic structure with randomly placed aluminium and magnesium was used for alloy study. The training convergence towards the number of structures trained on was tested, and $10^5$ grid points from each structure were chosen for training, as shown in **FIG. 5(a)**. The MAE approaches $8.0\times10^{-4}$ $e/\text{Å}^3$ with more than 5 structures of training data. **FIGs. 5(b)** and **(c)** indicate that the increase in system size barely influences the predictability of the machine learning model. The MAE in the system with 256 atoms is $8.14\times10^{-4}$ $e/\text{Å}^3$, slightly smaller than $7.66\times10^{-4}$ $e/\text{Å}^3$ from the system with 108 atoms. The detailed charge density is shown in the first row of **FIG. 6**. The absolute errors are mostly contributed from the center area of magnesium atoms, with a value of around $0.06$ $e/\text{Å}^3$.

Next, amorphous silicon and water reveal similar results, as shown in **FIGs. 5(d)-(i) and FIG. 6**. For amorphous silicon, the MAE reached below $0.004$ $e/\text{Å}^3$ after training with 7 structures of data. The max absolute error for the test structure is $0.037$ $e/\text{Å}^3$, nearing the silicon atoms. In the case of water, the MAE decreases below $0.0023$ $e/\text{Å}^3$ after training with 6 structures. The predictions in silicon with 216 atoms and water with 384 atoms are as good as that in smaller systems. However, with the increase in system size, the accuracy for water decreases relatively large, which may be induced by the rather small training structures. These results show that the accuracy of Deep Charge in amorphous structures is relatively lower than that of crystalline but can still be a good representation. Besides, the larger error around atom centers could be related to the cutoff function we used, which increases the weight of adjacent atoms but probably not be the best description of the electron distributions. A reasonable design of the cutoff function by reference to the near-core charge density should improve the predictability of machine learning models. The extremely high scalability will pave the way for predicting charge density in large-scale simulations.

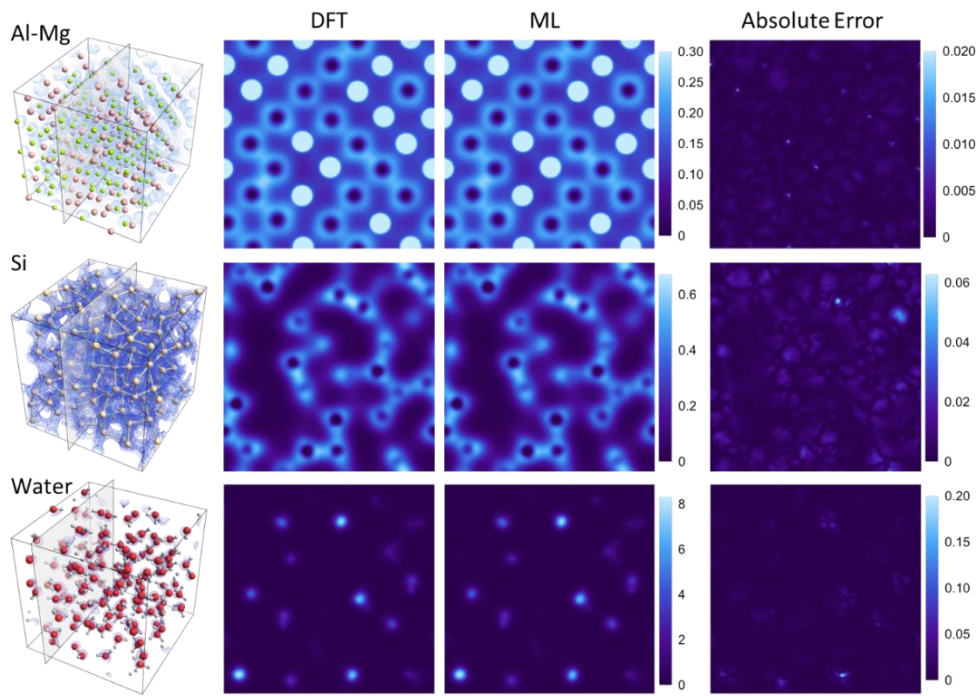

**FIG. 6.** Charge density in large aluminium-magnesium alloy, amorphous silicon, and water supercells. Noticed that the maximum value of charge density for aluminium-magnesium alloy is above 15.3 e/Å$^3$. We only display the charge density values in the interval of 0~0.3 e/Å$^3$ to show the details clearly.

## Conclusion

In summary, we have realized a new charge density representation with machine learning techniques. Our Deep Charge model can accurately represent charge density in semiconductor, metal, and molecular systems. The symmetry-preserving descriptor enhances the training efficiency. Only one single DFT calculation is enough for training extremely high accuracy models for simple crystal structures. It can also capture the fluctuations of the environment and represent complex systems by training with fewer data. Meanwhile, further improvement in the accuracy and efficiency of this model is still possible. A reasonable design of the cutoff function and loss function may increase the charge density accuracy near the atoms. Moreover, implementing a concurrent learning strategy can automate the sampling process of data points used for training, leading to more efficient learning and less human intervention. Our model is easy to implement and can provide an approach for fast charge density prediction avoiding the usage of KS quasi-particle wavefunctions. Looking forward, with the predicted charge density, many other physical quantities like energy and force can be derived theoretically based on the orbital-free DFT frameworks[23], which thus assist the electronic structure analysis, stability prediction, elastic behavior, etc. in large-scale condensed matter systems with the first-principal accuracy.

# Methodology

**Density functional theory calculations**

The charge density data are all based on the DFT calculations using the Vienna Ab initio Simulation Package (VASP)[3]. We used the Perdew-Burke-Ernzerhof (PBE)[24] exchange and correlation functional with PAW pseudopotentials to handle the core electrons[25]. The charge density data for training only contains the valence electrons. Explicitly, the electrons in 3s and 3p orbitals for silicon and aluminium, 2p and 3s orbitals for magnesium, 2s and 2p orbitals for oxygen were treated as valance electrons. Our models were then based on the charge density from valance electrons.

The electronic self-consistent calculation stops when the difference between the current and previous iteration is less than $10^{-6}$ eV for all the single-point calculations. For the accurate charge density calculation of silicon, the tetrahedron method with Blöchl corrections was adopted, with a 2×2×2 k-point mesh and a plane wave energy cutoff of 520 eV. A relatively low criterion was adopted for AIMD and geometry optimization, in which we set Gaussian smearing with a sigma value of 0.05, the electronic stop condition as 0.001 eV, and the ionic stop condition as -0.08 eV/Å. The timestep for the AIMD simulation was set to 3 fs. For aluminium, the first-order Methfessel-Paxton method was used for smearing, with a sigma of 0.22. The energy cutoff was 650 and Monkhorst-Pack k-point mesh with a k-spacing of 0.1 Å$^{-1}$ was adopted. The AIMD simulations of aluminium were the same as that of silicon but with a timestep of 5 fs. The settings for aluminium-magnesium alloy are similar to that of aluminium, but with a k-spacing of 0.2 Å$^{-1}$ for faster calculation. For $H_2O$, the SCAN functional[26] was used to model the exchange-correlation energy. The energy cutoff was set to 680 eV and the k-point spacing is 0.5 Å$^{-1}$.

**Datasets construction**

The structures for constructing datasets were sampled from AIMD or deep potential molecular dynamics (DPMD). We used AIMD under 300 K for crystalline silicon and aluminium. In amorphous silicon, we first ran an AIMD under 2000 K to obtain the randomly distributed atoms. Then geometry optimizations were performed to obtain the amorphous structures. On the other hand, we adopted DPMD with a Large-scale Atomic/Molecular Massively Parallel Simulator (LAMMPS)[27] for $H_2O$ and aluminium-magnesium alloy, for which the Deep Charge models were available from previous research[28,29]. The NPT simulations were under 300 K and 100 kPa. The initial structures of aluminium-magnesium alloy were generated by randomly replacing half of the aluminium atoms with magnesium in the face-centered cubic lattice. Then the randomly generated structures were relaxed for 0.5 ps with a timestep of 5 fs. For water, the timestep was set to 1 fs, and the structures were collected after simulating for more than 5 ps. The charge density data used in this work were then constructed from the obtained structures with DFT calculations described above. We collected the charge density generated from VASP. In total, we have collected 7 data frames for crystalline silicon, 109 data frames for amorphous silicon, 20 for aluminium, 20 for aluminium-magnesium alloy, and 10 for water. The final frame of data in each system was selected to be shown in this work as the testing structure. For each charge density file, we randomly chose $10^5$ data points as default, except when testing the convergence of accuracy towards the number of data points. These data points are separated into 500 batches, each batch contains 200 points for training or validating.

**Deep Charge model**

To realize the training of the Deep Charge model, it is important to notice that our grid point strategy can be equivalent to the usage of pseudo-atoms in the DP scheme. The charge density at a specific grid point can be regarded as the "atomic energy" of that point, which can be represented by the DP model[30]. Thus, the Deep Charge model can output the distribution by inputting the real atoms in the system together with the pseudo-atoms on each grid point and then collecting the predicted scalar values of the pseudo-atoms. The $r_c$ for all the models was set to 6.5 Å with an $r_s$ of 0.0 Å, where the value of $r_s$ did hardly affect the result. For H$_2$O, the training was operated for $2\times10^6$ steps, with a learning rate decreasing from $10^{-3}$ to $10^{-8}$ exponentially. We have chosen a three-layer ResNet structure of the form [20, 40, 80] for the embedding network and [200, 200, 200] for the fitting network. The same parameters are adopted for silicon, except for a longer training with $5\times10^6$ steps. For aluminium, the training was operated for $4\times10^6$ steps, with a learning rate decreasing from $10^{-3}$ to $10^{-7}$ exponentially. For aluminium-magnesium alloy, the training was operated for $3\times10^6$ steps, with a learning rate decreasing from $10^{-3}$ to $10^{-6}$ exponentially.

# Acknowledgments

We appreciate the discussion and coding instruction with Mr. Xuejian Qin and Mr. Yifan Shan. We greatly thank the information and suggestions provided by Dr. Linfeng Zhang. This work is partially funded by the Grand Challenge scheme of the Physics Foundation at the University of Sydney. The research undertakes resources provided by National Computational Infrastructure (NCI Australia) and the Pawsey supercomputer center. We acknowledge the financial support from the Australian Research Council (Grant No. DP200100940 and No. DE180100167), and National Key R&D Program of China (Grants No. 2021YFA0718900 and No. 2022YFA1403000).

# Supplementary Materials

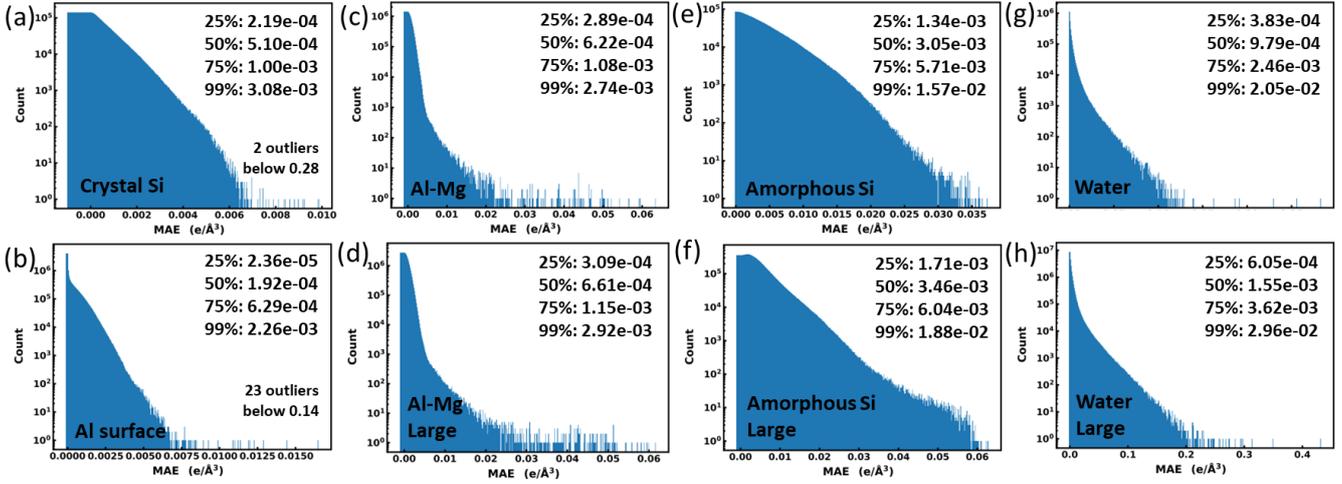

**FIG. S1**. The absolute error distribution for grid points in the tested systems. There are two outliers not drawn located at 0.277 and 0.281 e/Å$^3$ in the crystalline silicon, and 23 outliers in aluminium with values below 0.14 e/Å$^3$. The text in each chart indicates how many percent of the errors are below a certain value.

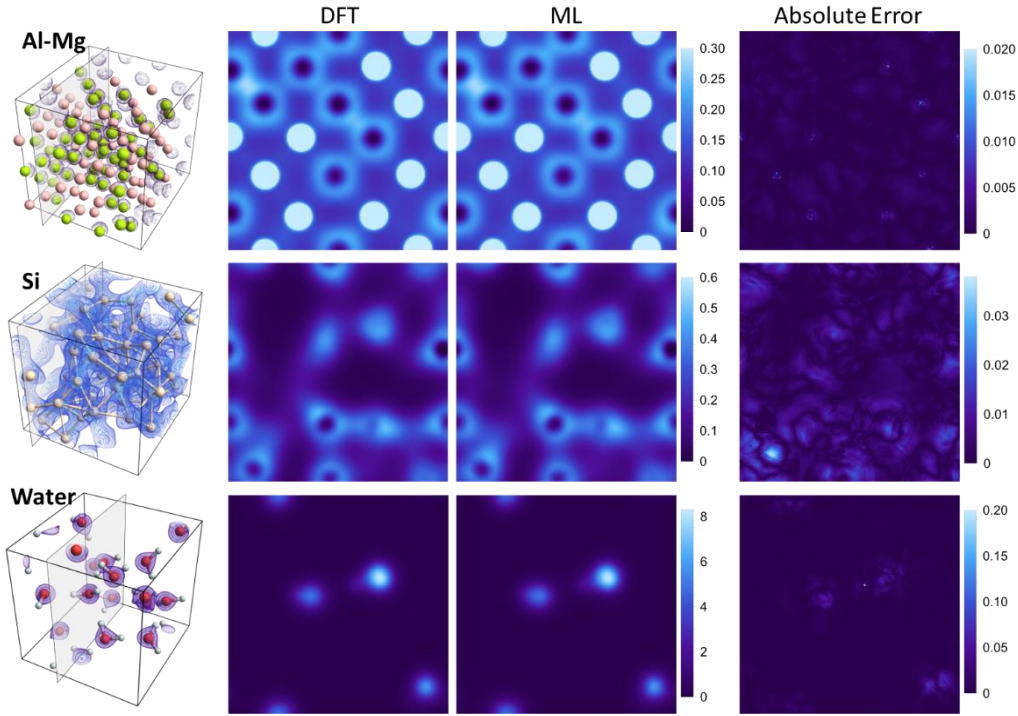

**FIG. S2.** From top to bottom shows the comparison of charge density computed from DFT and Deep Charge in small systems of aluminium-magnesium alloy, amorphous silicon, and water, respectively. The first column shows the testing structures. The middle two columns display the charge density slices obtained from DFT and machine learning within the plane with max absolute error, respectively. Noticed that the maximum value of charge density for aluminium-magnesium alloy is above 15.3 e/Å$^3$. We only display the charge density values in the interval of 0~0.3 e/Å$^3$ to show the details clearly. The fourth column represents the absolute errors.